\documentclass[twocolumn,aps,prc,floatfix]{revtex4}
\usepackage{graphicx}
\usepackage{amsmath}
\usepackage{dcolumn}
\usepackage{graphicx}
\usepackage{bm}
\usepackage{url}

\begin{document}

\title{A possible probe of the new fifth force}

\author{Gao-Chan Yong}
\affiliation{%
{Institute of Modern Physics, Chinese Academy of Sciences, Lanzhou
730000, China}
}%

\begin{abstract}

The proposed protophobic fifth force in \emph{Phys. Rev. Lett. {\bf \emph{117}}, 071803 (2016)} recently attracts much attention. To confirm or refute the existence of this new interaction, here I propose a method to probe the protophobic fifth force in dense matter that formed in nucleus-nucleus collisions. As expected, the protophobic fifth force has negligible effects on the usual observables in nucleus-nucleus collisions. While the protophobic fifth force evidently affects the value of final positively and negatively charged pions ratio at very high kinetic energies. The signal thus could be used to probe the protophobic fifth force in dense matter by nucleus-nucleus collisions' experiments at current experimental equipments worldwide.

\end{abstract}

\maketitle

\section{Introduction}

It is known to all that there are four fundamental forces, i.e., gravity, electromagnetism and the strong and weak nuclear forces. The electromagnetic force, a type of physical interaction that occurs between electrically charged particles,
is the one responsible for practically all the phenomena
one encounters in daily life above the nuclear scale,
with the exception of gravity.
The gravitation sculpts the universe at the enormous scale of galaxy clusters and the strong and weak nuclear forces prevail in the tiny interactions between subatomic particles. The four forces generally govern the interactions between all the matter in the universe. The Non-Newtonian gravity was suggested by Fischbach
\emph{et al} \cite{fisch86} and then was further studied using string theory or M theory \cite{Sushkov2011,Yang2012}.
The enunciation of non-Newtonian gravity hypothesis spawned a large
number of important experimental and theoretical results
\cite{tower88,nature92,Moody93,Hoyle01,Chia03,Kap07,Ade07,Sushkov2011,Yang2012}.
In the simplest models, non-Newtonian gravity can be characterized effectively by adding
a Yukawa term to the normal gravitational potential
through exchanging a boson \cite{Sushkov2011,Yang2012,Fis99,Adel03,fu71,Uzan03,wen09,xu13}.
However, up to now, people still have not found existing evidence of such
non-Newtonian gravity.

The standard model describes the fundamental subatomic particles that are the building blocks of all matter. While searching for new forces has growing interest mainly because of the inability of the standard model of particle physics to explain dark matter. Also the standard model can not explain why dark energy is causing the universe to expand at an increasingly faster rate.

Scientists have historically proposed various exotic particles as fifth-force carriers \cite{fayet09,Boe-prl,Kri09,Boe04,Jean03,Zhu07,Gninenko1201,Gninenko1201b,reece09,yong13eta}.
Recently, the experimental studies of decays of an excited state of $^{8}$Be to its
ground state proposed a possible new boson with mass about 17 MeV \cite{Krasznahorkay2016}. After looking for properties consistent with other experimental results, Feng and colleagues concluded that this new particle could be a ``protophobic X boson''. Such a particle would carry a protophobic fifth-force \cite{fengfifth2016,fengfifth20162},
which is an extremely short-range force that acts over distances only several times the width of an atomic nucleus. This force is a sort of analogue to electromagnetism, except where
electromagnetism acts on electrons and protons and ignores neutrons, this fifth force works between electrons and neutrons and ignores protons.

The fifth force might help scientists achieve the Grand Unified Theory. The fifth force might also be an entryway into understanding dark matter, mysterious particles making up the bulk of the mass of the universe that have yet to be observed.

Nowadays the possible discovery of the unknown fifth force of nature attracts much attention and causes much debate \cite{news2016}. To confirm or refute the existence of this dark force, here I propose a method to probe the protophobic fifth force in dense matter that formed in nucleus-nucleus collisions. The results show that the protophobic fifth force generally has negligible effects on the final observables in nucleus-nucleus collisions. While the protophobic fifth force evidently affects the value of final positively and negatively charged pions ratio at very high kinetic energies. The charged pion ratio at very high kinetic energies thus could be used to probe the protophobic fifth force in dense matter formed in nucleus-nucleus collisions.

\section{The methodology}

The study is based on the semiclassical Boltzmann-Uehling-Uhlenbeck (BUU) transport model \cite{bertsch}, which is quite successful in describing
dynamical evolution of nuclear reaction.
The BUU transport model describes time
evolution of the single particle phase space distribution function
$f(\vec{r},\vec{p},t)$, which reads
\begin{equation}
\frac{\partial f}{\partial
t}+\nabla_{\vec{p}}E\cdot\nabla_{\vec{r}}f-\nabla_{\vec{r}}E\cdot\nabla_{\vec{p}}f=I_{c}.
\label{IBUU}
\end{equation}
The phase space distribution function $f(\vec{r},\vec{p},t)$
denotes the probability of finding a particle at time $t$ with
momentum $\vec{p}$ at position $\vec{r}$. The left-hand side of
Eq.~(\ref{IBUU}) denotes the time evolution of particle phase
space distribution function due to its transport and mean field,
and the right-hand side collision item $I_{c}$ accounts for the
modification of phase space distribution function by elastic and
inelastic two body collisions.
$E$ is particle's total energy, which is equal to
kinetic energy $E_{kin}$ plus its average potential energy $U$.
While the mean-field potential $U$ of the particle is given
self-consistently by its phase space distribution function.

%\begin{figure}[th]
%%\centering
%\includegraphics[width=0.55\textwidth]{fig11.eps}
%\caption{(Color online) The strength of protophobic fifth force between two neutrons as a %function of distance.} \label{fifthf}
%\end{figure}
%
In the model, I use a 17 MeV mass gauge boson \cite{Krasznahorkay2016,fengfifth2016}, mediating a Yukawa force with range 12 fm. The strength of the coupling for $u$ quark is $-3.5 \times 10^{-3} e$, for $d$ quark is $7 \times 10^{-3} e$, where $e = 0.303$ is the usual electromagnetic coupling constant \cite{fengfifth2016}.
The fifth force between two particles is then given by \cite{fengprivate}
\begin{eqnarray}
F^{ij}_{fifth}= 1.44\times 10^{-3}c_{i} c_{j}(\frac{1}{r^{2}}+\frac{1}{\lambda r})e^{-r/\lambda},
\label{strengthf}
\end{eqnarray}
where $\lambda$ = 12 fm, $c_{i}$ and $c_{j}$ are the strength parameters of coupling of two interacting particles and $r$ is their distance.
The strength parameters of different particles used in Eq.~(\ref{strengthf}) are shown in Table \ref{notef}.
\begin{table}[th]
\caption{The Strength parameters used in Eq.~(\ref{strengthf}).}
\label{notef}%
\begin{tabular}{|c|c|}
  \hline
  particle $i$  & strength parameter $c_{i}/e$\\
  \hline
  $p$                    &    0       \\
  \hline
  $n$                    &  $1.05\times10^{-2}$    \\
  \hline
  $\pi^{-}$              & $1.05\times10^{-2}$     \\
  \hline
  $\pi^{0}$              & 0     \\
  \hline
  $\pi^{+}$              & $-1.05\times10^{-2}$   \\
  \hline
  $\Delta^{-}$           & $2.1\times10^{-2}$     \\
  \hline
  $\Delta^{0}$           & $1.05\times10^{-2}$     \\
  \hline
  $\Delta^{+}$           & 0     \\
  \hline
  $\Delta^{++}$          & $-1.05\times10^{-2}$     \\
  \hline
\end{tabular}
\end{table}

%Fig.\ \ref{fifthf} is an example of the strength of the fifth force between neutron and %neutron.
%
From Eq.~(\ref{strengthf}), it is seen that the strength of this fifth force decreases as increase of the distance of two particles and there is a sharp increase at short distance. The force is repulsive between two neutrons. From Eq.~(\ref{strengthf}) and Table \ref{notef}, it is shown that there is no fifth force between two protons. And proton, $\pi^{0}$, $\Delta^{+}$ have no fifth force interactions with other particles. More interestingly, $\pi^{-}$ and $\pi^{+}$ have opposite fifth force interactions with other particles.

\section{Results and discussions}

The fifth force should have less effects on general observables in heavy-ion collisions
but may have effects on light meson production. Because $\pi^{-}$ and $\pi^{+}$ have opposite fifth force interactions with other particles, to probe the fifth force in dense matter that formed in heavy-ion collisions, it is naturally to think of the observable $\pi^{-}/\pi^{+}$ ratio.
\begin{figure}[th]
\centering
\includegraphics[width=0.5\textwidth]{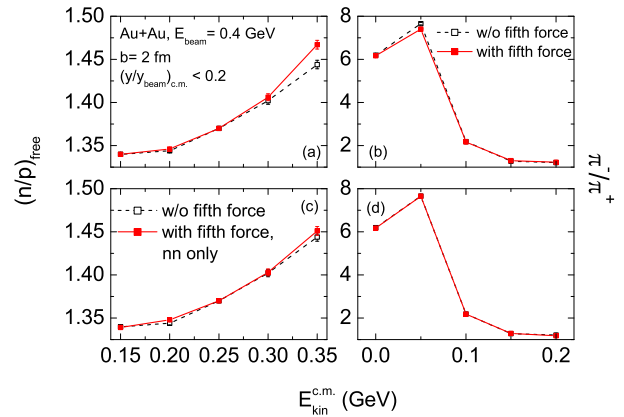}
\caption{(Color online) Effects of the fifth force on the kinetic energy distributions of the neutron to proton ratio of free nucleons and the $\pi^{-}/\pi^{+}$ ratio in the Au + Au reaction at 0.4 GeV/nucleon (in center-of-mass system). The cases in panels (a) and (b) consider the fifth force interactions among all kinds of particles in heavy-ion collisions, including neutron, $\pi$ meson and $\Delta$ resonances with different charge states. While the cases in panels (c) and (d) only consider the fifth force interactions between neutrons.} \label{kineticfifth}
\end{figure}
\begin{figure}[th]
\centering
\includegraphics[width=0.5\textwidth]{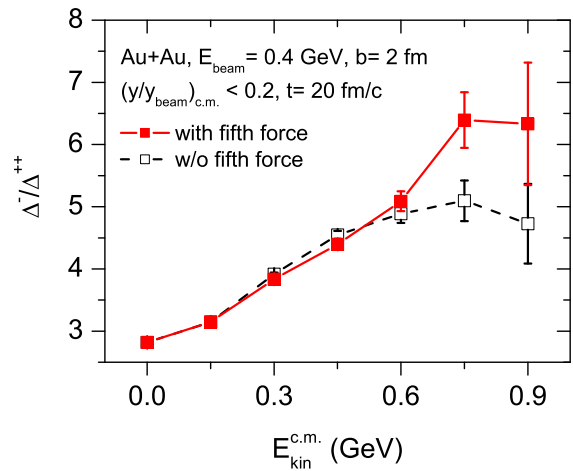}
\caption{(Color online) Effects of the full fifth force interactions on the kinetic energy distribution of $\Delta^{-}/\Delta^{++}$ ratio at maximal compression stage of the central Au + Au collision at 0.4 GeV/nucleon incident beam energy.} \label{deltaekin}
\end{figure}
Since the strength of the fifth force is large at short distance, it is naturally to think that the emitting particles from the compression region in heavy-ion collisions may carry more information about the fifth force. Fig.\ \ref{kineticfifth} shows kinetic energy distribution of the emitting neutron to proton ratio of free nucleons $(n/p)_{free}$ and the $\pi^{-}/\pi^{+}$ ratio in central Au + Au reaction at 0.4 GeV/nucleon. From panel (a), it is seen that due to the repulsive fifth force interactions of neutrons, more neutrons are emitted from dense matter formed in heavy-ion collisions thus more free neutrons are produced. Therefore the fifth force increases the value of free $n/p$ ratio as shown in panel (a) in Fig.\ \ref{kineticfifth}. The right upper panel of Fig.\ \ref{kineticfifth} shows the corresponding $\pi^{-}/\pi^{+}$ ratio in the same Au + Au reaction at 0.4 GeV/nucleon. Due to more free neutron emissions, the dense matter formed in heavy-ion collisions is neutron-deficient. Less neutron-neutron collisions cause less $\pi^{-}$ productions \cite{bali2005}, thus the value of the $\pi^{-}/\pi^{+}$ ratio decreases with the fifth force interactions as shown in the right upper panel (b) of Fig.\ \ref{kineticfifth}. As comparison, the cases in panels (c) and (d) of Fig.\ \ref{kineticfifth} only consider the fifth force interactions among neutrons, i.e., turning off the fifth force interactions of $\Delta$ resonances and $\pi$ mesons. It is seen that the effects of the fifth force become smaller comparing to that with full fifth force consideration (the full fifth force considers all the fifth force interactions among all kinds of particles) as shown in panels (a) and (b). One thus can deduce that the behavior of resonance dynamics affects the free neutron to proton ratio and the $\pi^{-}/\pi^{+}$ ratio at relative low kinetic energies.

In heavy-ion collisions, relative numbers of very energetic neutrons (protons) emitted from dense matter are mainly affected by the relative numbers of produced energetic $\Delta^{-}$ ($\Delta^{++}$) in dense matter ($\Delta^{-}$ mainly decays into $\pi^{-}$ and neutron and $\Delta^{++}$ mainly decays into $\pi^{+}$ and proton). From Table \ref{notef}, it is seen that $\Delta^{-}$'s suffer stronger repulsive fifth force interactions than neutrons while $\Delta^{++}$'s suffer attractive fifth force interactions from matter, thus there should be more energetic $\Delta^{-}$ productions than $\Delta^{++}$. Fig.\ \ref{deltaekin} shows the effects of the fifth force on the $\Delta^{-}/\Delta^{++}$ ratio in dense matter as a function of kinetic energy. It is clearly shown that the value of the $\Delta^{-}/\Delta^{++}$ ratio increases with the fifth force at very high kinetic energy, i.e., there are relative more $\Delta^{-}$ productions than $\Delta^{++}$ at very high kinetic energy. This is the reason why the effects of the fifth force interactions are so small with only neutron-neutron fifth force interactions as shown in panels (c) and (d) of Fig.\ \ref{kineticfifth}. From the upper panels (a) and (b) of Fig.\ \ref{kineticfifth}, it is seen that maximal effects of the fifth force on the free n/p ratio and the $\pi^{-}/\pi^{+}$ ratio in central Au + Au reaction at 0.4 GeV/nucleon are only about 1.5\% and 3.5\%, respectively. However, since Fig.\ \ref{deltaekin} shows that the energetic $\Delta^{-}/\Delta^{++}$ ratio is more sensitive to the fifth force, the energetic $\pi^{-}/\pi^{+}$ ratio should be also more sensitive to the fifth force. This is because the ratio of $\pi^{-}$ and $\pi^{+}$ yields are mainly decided by the relative numbers of the $\Delta^{-}$ and $\Delta^{++}$ resonances \cite{bali2005}.

\begin{figure}[th]
\centering
\includegraphics[width=0.5\textwidth]{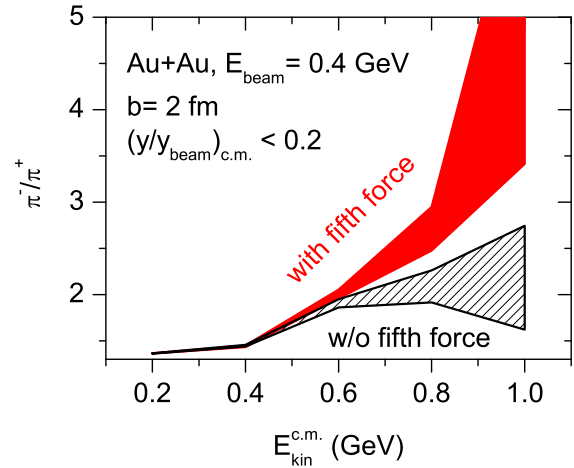}
\caption{(Color online) Effects of the fifth force on the energetic $\pi^{-}/\pi^{+}$ ratio in the Au + Au reaction at 0.4 GeV/nucleon. The fifth force interactions are from all kinds of particles.} \label{energeticfifth}
\end{figure}
To show more clearly the effects of the fifth force on the $\pi^{-}/\pi^{+}$ ratio, in Fig.\ \ref{energeticfifth}, we plotted kinetic energy distribution of the $\pi^{-}/\pi^{+}$ ratio at very high kinetic energies. In the simulations, we accumulated 5,000,000 events for each case. Energetic $\pi^{-}$ and $\pi^{+}$ mesons emitted from the dense matter are mainly from energetic nucleon-nucleon collisions (neutron-neutron collision mainly forms $\Delta^{-}$ resonance and it decays into neutron and $\pi^{-}$ meson and proton-proton collision mainly forms $\Delta^{++}$ resonance and it decays into proton and $\pi^{+}$ meson) in the compression stage in heavy-ion collisions. Because the fifth force repels $\Delta^{-}$'s and attracts $\Delta^{++}$'s (as shown in Table \ref{notef} and Fig.\ \ref{deltaekin}), more $\Delta^{-}$'s are produced than $\Delta^{++}$'s at very high kinetic energies. It is thus not surprising that one sees the fifth force increases the value of the $\pi^{-}/\pi^{+}$ ratio at very high kinetic energies as shown in Fig.\ \ref{energeticfifth}.

The production of very energetic pions in heavy-ion collisions at intermediate energies is definitely from many rounds' nucleon-nucleon or baryon-baryon scatterings, thus both the colliding energy and the effects of the fifth force are accumulated.
From Fig.\ \ref{energeticfifth}, it shows that the maximal effect of the fifth force on the final $\pi^{-}/\pi^{+}$ ratio is larger than 25\%.
The uncertainty band denotes the statistical error.
The number 25\% is deduced from the lower band of the value of $\pi^{-}/\pi^{+}$ ratio with the fifth force (3.5) and the upper band of the
value of the $\pi^{-}/\pi^{+}$ ratio without the fifth force (2.8).
It is worth mentioning that such effect could be measured at current experimental equipments worldwide, such as heavy-ion collisions at RIBF-RIKEN in Japan \cite{shan15}, at CSR in China \cite{ZGX09,csr} or at GSI in Germany \cite{GSI}. In the study, each event produces about $3\times10^-5$ pions with kinetic energy of about 0.8 GeV in Au + Au reaction at 400 MeV/nucleon. One thus needs to accumulate at least millions of events.

Because the strength of the fifth force is very small, the collision of two \emph{heavy} nuclei is best used to probe the fifth force. The heavy collision system can get stronger effects of the fifth force due to the superposition of fifth forces among more particles. And also the heavy collision system can form larger dense matter size thus more rounds of nucleon-nucleon or baryon-baryon scatterings accumulate more effects of the fifth force in nucleus-nucleus collision. With the increase of the incident beam energy of nucleus-nucleus collisions, the effect of the fifth force on the energetic positively and negatively charged pions ratio is negligible as the thermal or random energies also become larger.

While the nuclear symmetry energy may affect the value of $\pi^{-}/\pi^{+}$ ratio in heavy-ion collisions at intermediate energies \cite{bali2005}. For very energetic $\pi^{-}/\pi^{+}$ ratio, however, the symmetry energy should show no effects. This is because after many rounds's nucleon-nucleon or baryon-baryon scatterings, the isospin effects on the final energetic nucleon-nucleon or baryon-baryon collisions are averaged and thus planished. The produced energetic pions are therefore not affected by the exact form of the symmetry energy. Similarly, the isospin-dependence of the in-medium baryon-baryon cross section in the above energetic collisions is also planished due to many rounds's different isospin-dependent baryon-baryon scatterings. Since there are no effects of the symmetry energy on the produced energetic nucleons and resonances, the threshold effect on energetic pion production can be also neglected \cite{ko2017}. Of course, there are also some others theoretical uncertainties which affect the value of energetic charged pion ratio, such as initialization of nucleon distribution in coordinates and momentum space in projectile and target nuclei, dynamical self-energies due to nucleon-nucleon short-range correlations and core polarization \cite{zhang2016}, off-shell pion transport \cite{cassing2000}, etc. However, all the above effects on energetic charged pion ratio are negligible \cite{zhang2016,cassing2000}.

There are still other mechanisms which cause the enhancement of energetic $\pi^{-}/\pi^{+}$ ratio with the fifth force in heavy-ion collisions at intermediate energies.
For example, based on the $\Delta$-hole model and the relativistic Vlasov-Uehling-Uhlenbeck transport model with the relativistic nonlinear NL$\rho$ interaction, the pion $p$-wave potential may enhance the value of the energetic $\pi^{-}/\pi^{+}$ ratio at pion kinetic energies above 200 MeV \cite{ko2017}, although the pion potential may have no effects at pion kinetic energies around 200 MeV \cite{guoyzl2015}. Nevertheless, the evident enhancement of energetic $\pi^{-}/\pi^{+}$ ratio at pion kinetic energies around 900 MeV are mostly due to the fifth force because the very energetic pions almost have no much time to suffer the mean-field potential from surrounding nuclear matter.

\section{Conclusions}

Once the recently proposed new fifth force is confirmed by experiments, it would have profound influence in fundamental physics. To confirm or refute this kind of new interaction, I propose to probe the fifth force in dense matter that formed in nucleus-nucleus collisions. It is shown that the protophobic fifth force affects the value of final positively and negatively charged pions ratio at very high kinetic energies in nucleus-nucleus collisions at intermediate energies. The \emph{heavy} collision system can effectively enlarge the effects of such dark force on the ratio of yields of positively and negatively charged pions. This signal could be used to probe the protophobic fifth force by nucleus-nucleus collisions' experiments at current experimental equipments worldwide.

\section*{Acknowledgements}

The author acknowledges Jonathan L. Feng for helpful communications. The work was carried out at National Supercomputer Center in Tianjin, and the calculations were performed on TianHe-1A. The work is supported by the National Natural Science Foundation of China under Grants No. 11775275, and No. 11435014.

\end{document}